\documentclass[10pt, twoside, a4paper]{article}
\usepackage{amssymb}
\usepackage[figuresright]{rotating}
\usepackage{color}
\usepackage{abstract}
\usepackage{MnSymbol}
\usepackage{graphicx}
\usepackage{multicol}

\begin{document}


\title{Forecasting the direction of incoming radiation based on fusion of gyroscopic and spectroscopic data\\ \vskip10pt}
\author{ Marcus J. Neuer, Christian Henke and Elmar Jacobs\\{\normalsize innoRIID GmbH., 41516 Grevenbroich, Germany}}

\maketitle
\begin{onecolabstract}
\bfseries
A method is shown to estimate the position of by fusing the data from a sodium iodide detector and a gyroscope while panning the detector. Based on simple geometry considerations, the search motion of a detector is modelled as angular panning. Correlation of both sensor streams is shown as a predictor for the direction of the incoming radiation. The method also allows for a rough distance classification, where a distant source or a homogenous radiation field can be distinguished from a single source near to the detector.  
\end{onecolabstract}
\vskip20pt

\section{Introduction}
Handheld radioisotope identification devices (RIID) are widely used as security equipment. They complement static portal detection philosophies and allow their users to actively search for sources by moving the device. Most instruments therefore contain an extra mode for this search to give visual, acoustic and vibrational feedback about the strength of the current radiation field. In the best case, they finally guide their users to the hotspot or source of radiation.

A very interesting feature would be a compass-like needle, pointing roughly towards the direction where a radiation hotspot is assumed. Please regard also to \cite{RisticGunatilaka:2008} and reference therein. There are different ways to do this. First, one can use multiple detectors, in the easiest configuration two. Then, common approaches as described in \cite{Neuer:2008} apply and the position of the source can be approximated even with a certain amount of perturbation material within the path of the incoming radiation. But there are also several reasons not to use more than one detector in an handheld instrument: (a) it is expensive, (b) it increases the complexity of the electronics system and (c) it increases the points of possible failures. 

More elaborated approaches also use multi-detector concepts such as coded-aperture applications \cite{Fenimore:1978-1,Fenimore:1978-2},\cite{Fenimore:1989} or even Compton-cameras. The latter have seen a significant evolution and with statistical algorithms as presented in \cite{Wilderman:1998}, such cameras can locate a source very reliable. Nevertheless, solutions like these two examples are not within our scope. We refer to a rather inexpensive handheld configuration with only one detector system. 

We therefore adopted a different approach based on sensor or data fusion \cite{Donati:2006}, whereas we explicitly refer to the merging of information from two different, heterogenous sources. In our case this is the nuclear detection system and a gyroscope (gyro) \cite{Heredia:2008}. Adding a gyro to the detection system is an easy task. Such modules are custom-out-of-the-shelf products and can be coupled to a lot of interesting hardware concepts, e.g. a Raspberry Pi micro computer platform. On the other hand, most mobile platforms like Android boards or the iPhone are already fully equipped with these sensors. Once the data of the gyro is analysed, we can reconstruct a part of the local detector motion. Together with the count rate given by the radiation detector we have enough data to predict the direction to the source. 

Data fusion has several prospects as it adds information for the interpretation of a given sensor. Nowadays, it is used in a wide range of applications \cite{Dasarathy:2003,Dasarathy:2007}. Most prominently though not stated exactly as data fusion, examples also include the application of sensor fault detection \cite{Khalastchi:2013}.

Our technique represents a virtual or so-called soft sensor. With that, the fused result of both sensorial inputs represents a new, algorithmically deduced sensor component that gives the answer to the questions from where the source radiation is coming from. 

Note that we do not involve any shielding of the detector, nor do we restrict to collimated source beams. Instead we want to establish a simple, robust but certainly rough mean to estimate the direction to the source.

Our presentation starts with some very basic models to gain insight into the setup. It is clear to us, that real field missions are practically unforeseeable. The multitude of different measurement geometries not only including scattering and absorption effects, but also distributed sources,  render it quasi impossible to calculate the radiation direction precisely. Nonetheless, we believe that a rough estimator, e.g. indication whether the source is left or right of the search path, is still achievable.

\begin{figure}[t]
\centering
\includegraphics[angle=0,width=0.8\columnwidth]{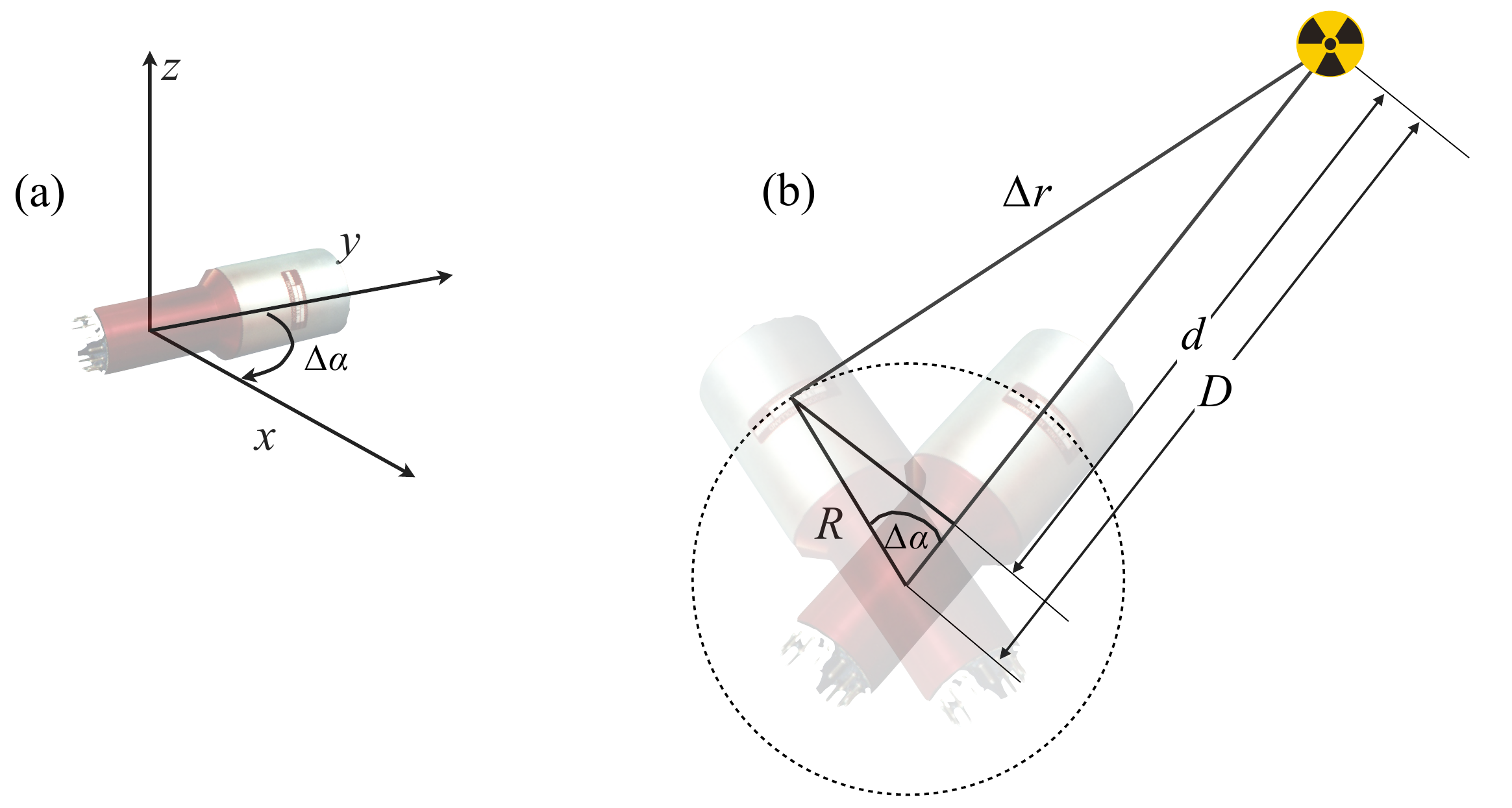}
\caption[Scheme]{Sketch of our geometry. (a) Side view of the detector and our axis definitions as well as the angle $\alpha$. (b) Top view. Distance of the rotation center to the source $D$.  The detector panning is assumed to be circular.}\label{Scheme}
\end{figure}

\section{Sensor fusion approach}
\subsection{Approximative relation between count rate and geometry}

Key idea is the hereby the knowledge of the gyro states, in detail knowledge of the local angular motion of our instruments. We record this motion in terms of three coordinates $\alpha, \beta, \gamma$, their respective velocities $v_\alpha=d\alpha/dt,\, v_\beta = d\beta/dt, \, v_\gamma = d\gamma/dt$ and accelerations $a_\alpha=d^2\alpha/dt^2,\, a_\beta = d^2\beta/dt^2, \, a_\gamma = d^2\gamma/dt^2$. Available gyro components grant experimental access to these quantities. For the sake of simplicity, we define the angle $\alpha$ to correspond to our motion. That means, we pan the instrument around the $z$-axis, within the $x-y$-plane.  

We apply a very simple circular model for the search motion of a handheld. Lets for a moment assume we know the distance $D$ between instrument rotation center and source. $R$ may be the radius of our search motion, $\alpha$ the angle reported by the gyroscope, with $d = D-R\cos{\alpha}$ it follows that

\begin{equation}
{\Delta r}^2 = (D-R\cos{\alpha})^2 + R^2\sin^2\alpha. \label{DeltaR}
\end{equation}

Here $x$ is now the resulting distance to the source during rotation. Please regard to Fig. \ref{Scheme} for further details. 

\subsection{Count rate}

In our model of the source we define the count rate at the sources physical surface to be $A_0$. Following the quadratic law, we can relate the distance between source and detector to the count rate \cite{RisticGunatilaka:2008},

\begin{equation}
A(\Delta r) = \frac{A_0}{({\Delta r} +1)^2}.
\end{equation}

and further to angle $\alpha$ from (\ref{DeltaR}),

\begin{equation}
A(\alpha) = \frac{A_0}{( (D-R\cos{\alpha})^2 + R^2\sin^2\alpha +1)^2}. \label{RotatingPattern}
\end{equation}

\subsection{Simplified models of detector motion}

\subsubsection{Rotation}

\begin{figure}[t]
\centering
\includegraphics[angle=0,width=1.0\columnwidth]{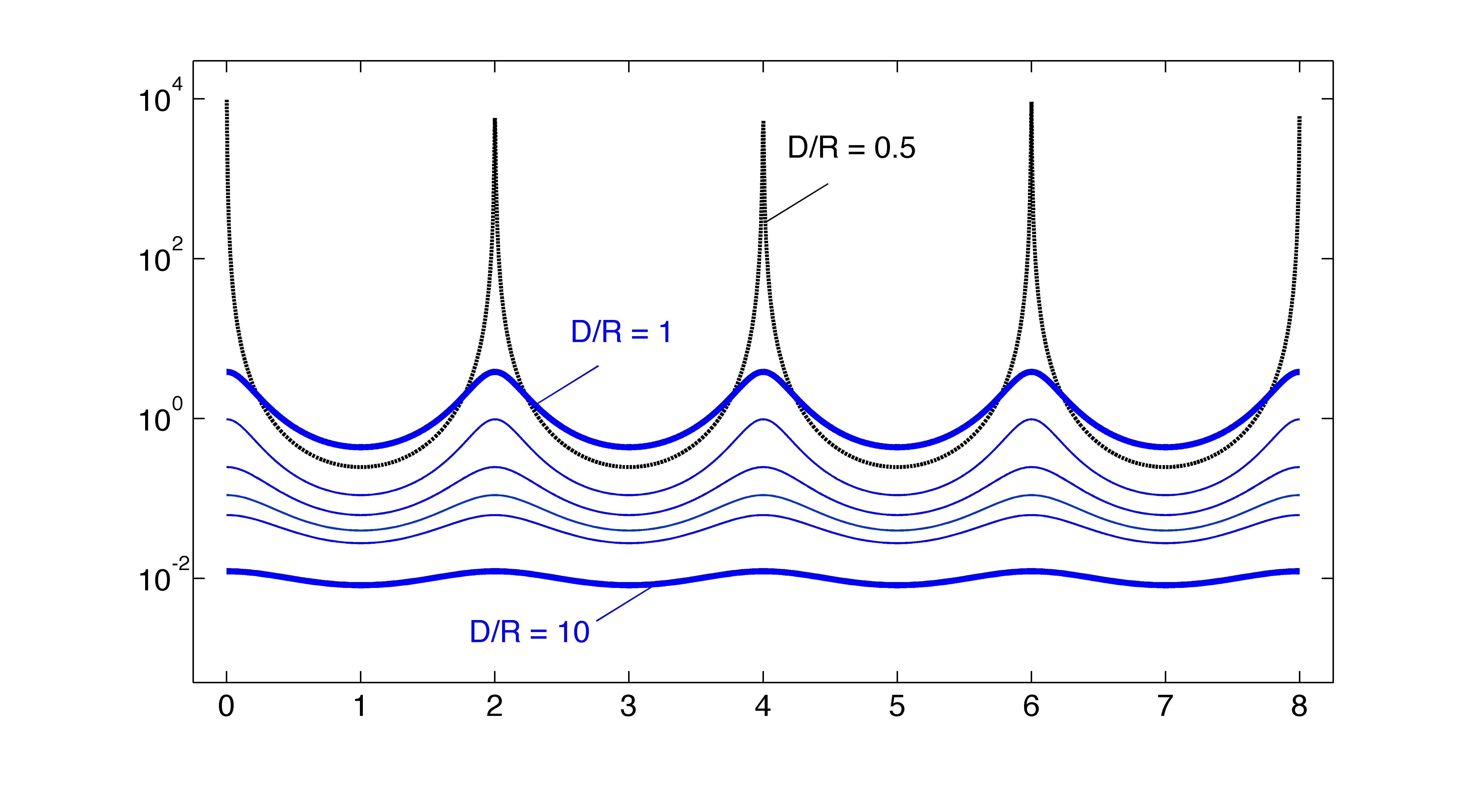}
\caption[Rotating]{Count rate $A$ as function of the angle $\alpha$, while the detector would be rotating. Different ratios $D/R$ are plotted.}\label{Rotating}
\end{figure}

We first start with a simple rotation. It can be easily checked whether the output of our gyroscope considerations is correct, once the 
In Fig. \ref{Rotating} we show an analytical evaluation of (\ref{RotatingPattern}), yielding our assumption on what the count rate at our detector should look like, if it is rotated along the $z$-axis within the radiation field. The result depends on the distance $D$ and the panning radius $R$. Here it is natural that with $D/R\rightarrow\infty$ the impact of rotating completely vanishes. 

\begin{figure}[t]
\centering
\includegraphics[angle=0,width=1.0\columnwidth]{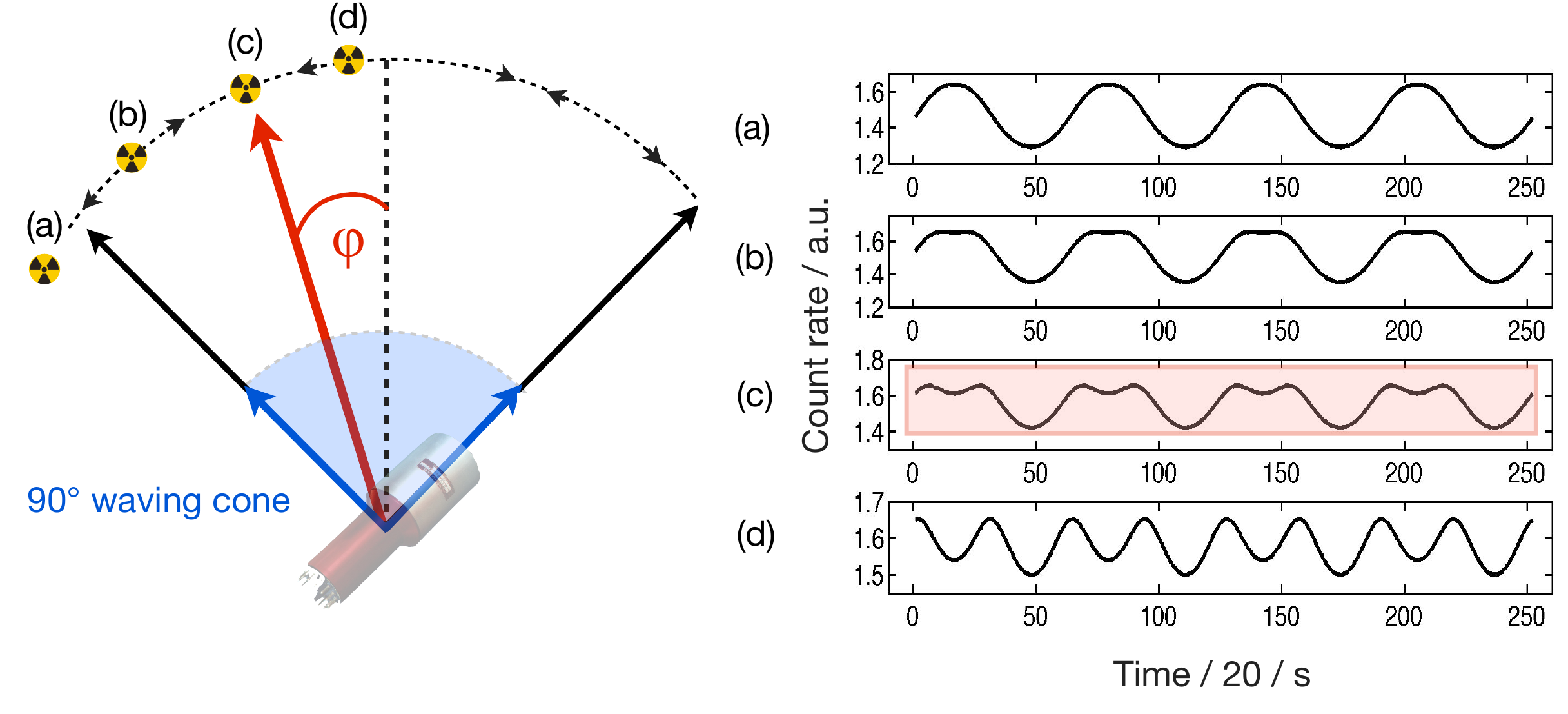}
\caption[Rotating]{Count rate $A$ as function of the angle $\alpha$, while the detector is panned for the ratio $D=2R$ and an approximate count rate of $A=1600\textrm{cps}$. For comparison with real sources, our detector has a sensitivity of about $S=2400\textrm{cps}/\mu \textrm{Sv}/\textrm{h}$ for cesium $^{137}\textrm{Cs}$. The simulation assumes a distance so that the dose rate becomes $\rho=1.6\mu \textrm{Sv}/\textrm{h}$.} \label{Waving}
\end{figure}

\subsubsection{Waving}

The most relevant search motion for finding a source is panning or waving, in other words moving the detector from right to left and back again. With the previous Eq. (\ref{RotatingPattern}) this behaviour can be described by restricting the circle to sort of waving cone. In right part of Fig. \ref{Waving} we show a sketch of this motion.

Following scenarios can be found with this motion and for each scenario Fig. \ref{Waving} depicts the expected count rate as function of time: (a) the source is at the left-(right)-hand-side of the detector. In this situation the count rate rises till we reach the point closest to the source. It would be a maximum, if the source is really in line with the final position of the detector. (b) the source within the $90^\circ$ waving cone, yielding an elongated maximum in the count rate. (c) the source is approximately at the middle of the left quarter of our waving motion. The count rate rises, has a first maximum when passing the source direction, decreases to an intermediate minimum, because the source is not as far away as on the other side and finally it passes the source direction again with the returning motion leading into the next oscillation. This curve and the associated source position are marked in red. As last case, (d) marks a situation where the incoming radiation is close to the center axis of our waving motion and no indication is possible. Here, the count rate yields an oscillation peaks with similar heights. Note that here, each panning motion passes the highest count rate twice, so there are also twice the number of peaks as in scenario (a).

\subsection{Forecasting based on the angular information}\label{rules}

\begin{figure}[t]
\centering
\includegraphics[angle=0,width=0.7\columnwidth]{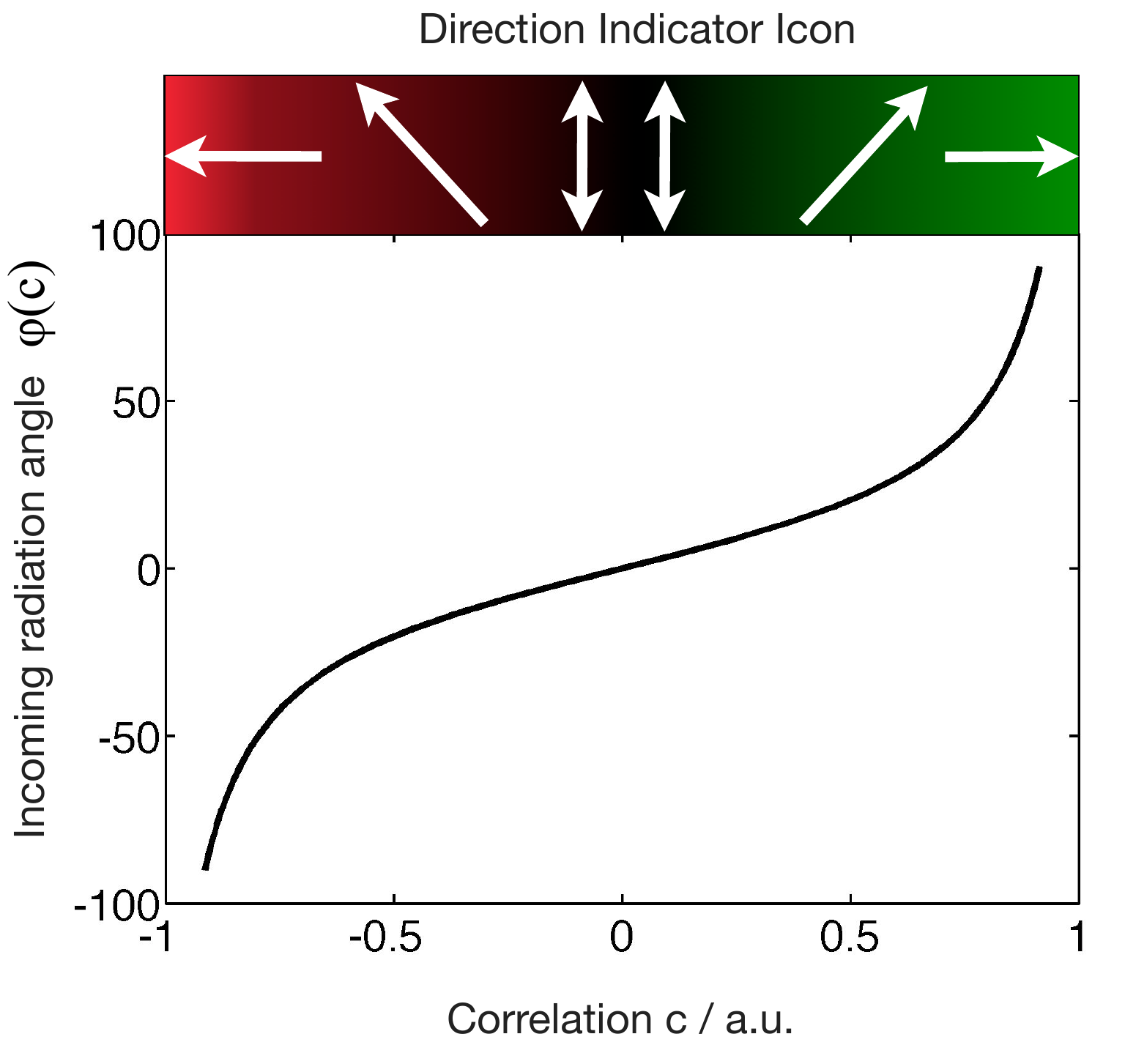}
\caption[Rotating]{Count rate $A$ as function of the angle $\alpha$, while the detector is panned for the ratio $D=2R$ and an approximate count rate of $A=1600\textrm{cps}$. For comparison with real sources, our detector has a sensitivity of $S=980\textrm{cps}/\mu \textrm{Sv}/\textrm{h}$ at the energy of the cesium $^{137}\textrm{Cs}$ photo peak at $661.65\textrm{keV}$. The simulation assumes a distance so that the dose rate becomes $\rho=1.6\mu \textrm{Sv}/\textrm{h}$.} \label{Correlation}
\end{figure}

Of course, the ratio $D/R$ has the same impact on the results as in the case with rotation. From the shapes found by panning the detector we constitute following data fusion rules:

\begin{itemize}
\item Once there is no change of count rate during panning, there are distributed sources or a single source far away. 
\item If the function of count rate yields significant changes, especially peaks, there is at least one source close by. 
\item The number of oscillation peaks is a first indication whether the incoming radiation is coming from the front or from left or right.   
\end{itemize}

and last, but perhaps our most important statement, 
\begin{itemize}
\item Let $\alpha(t)$ be the current local waving angle as function of the time $t$ and let $a(t)$ be the count rate of our detector, then the correlation value between both $C[a(t_S), \alpha(t_S)]$ is an directional measure for the incoming radiation. 
\end{itemize}

\section{Implementation in the prototypical and commercial scope}

\subsection{Prototype implementation}
A simple prototype was assembled using a Raspberry Pi micro-computer and an $I^2C$ based gyro module. A multi-channel analyser was mounted on a $3''\times1''$ sodium iodide detector and connected to the Raspberry Pi via serial communication. To verify our ruleset presented in \ref{rules}, we programmed a python software running on the Raspberry Pi, reading out the gyroscope and the MCA counts simultaneously. Using the prototype, the statements of \ref{rules} were confirmed and the algorithm was further improved with regard to performance. 

\subsection{Commercial exploitation}
To demonstrate the usability and robustness of the developed technique, we integrated a gyroscope into the hardware of a commercially available radioisotope identification device (RIID), which features a specific software user interface called "detect mode". This significantly extended the scope of our initial prototype. Figure \ref{Correlation} shows the algorithm behind the directionality. Depending on the determined value of $C$, a visual indication is given to the user of the instrument, in which direction the source is located. 
The instrument automatically identifies the motion type by its time pattern.

\footnotesize
\bibliographystyle{ieeetr}
\bibliography{./NuclearDetection}

\end{document}